\documentclass[preprint,aps,12pt]{revtex4}

\usepackage{graphicx}
\usepackage{dcolumn}
\usepackage{bm}
\usepackage{color}

\usepackage{amsmath}
\usepackage{amssymb}

\def\simle{\mathrel{\rlap{\raise 0.511ex \hbox{$<$}}{\lower 0.511ex \hbox{$\sim$}}}}
\def\simge{\mathrel{ \rlap{\raise 0.511ex \hbox{$>$}}{\lower 0.511ex \hbox{$\sim$}}}}

\newcommand \beq{\begin{eqnarray}}
\newcommand \eeq{\end{eqnarray}}
\def\simle{\mathrel{\rlap{\raise 0.511ex \hbox{$<$}}{\lower 0.511ex 
\hbox{$\sim$}}}}
\def\simge{\mathrel{ \rlap{\raise 0.511ex 
\hbox{$>$}}{\lower 0.511ex \hbox{$\sim$}}}}
\newcommand{\del}{\partial}

\newcommand{\kp}{\kappa}

\newcommand{\pk}{\partial_\kappa}

\begin{document}

\title{Exact renormalization group and $\Phi$-derivable approximations}

\author{Jean-Paul Blaizot}
\email{jean-paul.blaizot@cea.fr}
\affiliation{Institut de Physique Th\'eorique, CEA-Saclay, 91191 Gif-sur-Yvette Cedex, France.}
\author{Jan M. Pawlowski}
\email{pawlowski@thphys.uni-heidelberg.de}
\affiliation{Institut f\"ur Theoretische Physik, University of Heidelberg, Philosophenweg 16, 69120 Heidelberg, Germany.}
\author{Urko Reinosa}
\email{reinosa@cpht.polytechnique.fr}
\affiliation{Centre de Physique Th\'eorique, Ecole Polytechnique, CNRS, 91128 Palaiseau Cedex, France.}

\date{\today}

\begin{abstract}
We show that the so-called $\Phi$-derivable approximations can be combined with the exact renormalization group to provide efficient non-perturbative approximation schemes. On the one hand, the $\Phi$-derivable approximations allow for  a simple truncation of the infinite hierarchy of the renormalization group flow equations. On the other hand, the  flow equations turn the non linear equations that derive from the $\Phi$-derivable approximations into an initial value problem,  offering new practical ways to solve these equations. 
\end{abstract}

\maketitle

This paper deals with two non perturbative approaches to quantum field theory: the exact renormalization group (exact RG) (for reviews see  \cite{Berges:2000ew}) and the so-called  $\Phi$-derivable approximations \cite{LW,Baym}, also known as the two-particle-irreducible (2PI) effective action formalism \cite{Cornwall:vz}.  There has been recently a revival of interest in the application of the latter to various problems in field theory  including the thermodynamics of quantum fields \cite{Blaizot:1999ip}, the calculation of transport coefficients \cite{Aarts:2003bk} or the study of far from equilibrium quantum dynamics \cite{Berges:2004yj}. Of particular relevance to the present work are the recent studies of the renormalizability of $\Phi$-derivable approximations \cite{vanHees:2001ik}, and the specific schemes that were presented to implement this renormalization. On all these issues of renormalization, the exact RG can shed a new light   and this, we believe,  has not been exploited yet.  

The exact RG allows for the formulation of powerful non perturbative approximations. It leads generically  to an infinite hierarchy of coupled flow equations for the $n$-point functions, whose solution requires in practice some truncation.
The purpose of this paper is precisely  to show how one can exploit  the relations that exist between the $n$-point functions of  the 2PI  formalism in order to obtain a simple truncation of the RG flow equations. Note that the strategy presented in this  letter differs from that exposed in Ref.~\cite{Dupuis:2005ij} which exploits the RG invariance of the Luttinger-Ward functional. (Further discussion of the relations between the 2PI formalism and the exact RG can be found in the fifth reference  \cite{Berges:2000ew}.)
The truncation that we propose here is such that the  solution of the flow equation, for appropriate initial conditions,  coincides with an exact resummation of selected Feynman diagrams associated with the 2PI skeletons that are considered. Thus, in return, the truncated flow equations provide a powerful tool for  solving the 2PI equations: these are indeed  formulated  as an initial value problem, which is generally simpler to solve than the gap equations that naturally emerge in the 2PI approach. 

In this letter, we illustrate the proposed truncation on the simplest possible case, allowing for a concise presentation:  a massive $\varphi^4$ theory in $3$ dimensions, in the vacuum and in the symmetric phase.  Admittedly,  this super-renormalizable theory does not reveal the full power of the flow equations in dealing with some aspects of ultraviolet renormalization, such as for instance the intricacies of subdivergences.  These will be uncovered in the more detailed treatment of  renormalizable field theories in four dimensions, to be given in a forthcoming publication \cite{BPR}. Let us just mention here that the main results of the present  paper  generalize to 4-dimensional scalar theories, although the proofs are more involved in the latter case than in the 3-dimensional case that we consider here. \\

\begin{figure}[htbp]
\begin{center}
\includegraphics[width=8cm]{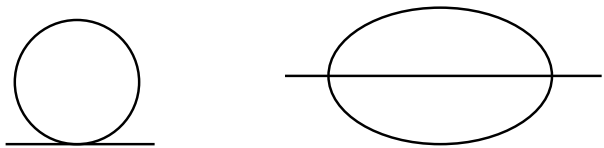}
\caption{``Tadpole'' and ``sunset'' diagrams that contribute to the 2-point function $\Gamma^{(2)}(p)$ and that are ultraviolet divergent. These two diagrams are also the skeletons that contribute to $\Sigma[G]$ respectively at two-loop and three-loop orders in the loop expansion of $\Phi[G]$. }\label{fig:example}
\end{center}
\end{figure}

The classical (Euclidean) action is written as
\beq\label{eactON} 
S[\varphi] =\int d^{3}x\,\left\lbrace{ \frac{1}{2}}\left(\del\varphi(x)\right)^2  + \frac{m^2_{\rm b}}{2} \, \varphi^2(x) + \frac{\lambda}{4!}\,\varphi^4(x) \right\rbrace.
\eeq
In the framework of  standard perturbative calculations this is to be considered as the ``bare'' action, with $m_{\rm b}$ the bare mass, and calculations are to be done in the presence of an ultraviolet  regulator characterized by a cut-off scale $\Lambda_{\rm uv}$. The specific form of this regulator will not be needed here. 
 Usual power counting reveals that, aside from the vacuum diagrams, only the 2-point function $\Gamma^{(2)}(p)$ is ultraviolet divergent. In fact, only two of the diagrams contributing to $\Gamma^{(2)}(p)$ are globally divergent. These are displayed in Fig.~\ref{fig:example}: The ``tadpole'' diagram is momentum independent and linearly divergent, while the ``sunset" diagram is momentum dependent but its logarithmic divergence is momentum independent. The divergences can then be ``absorbed'' in the bare mass, that is,  $m_{\rm  b}$ can  be adjusted as a function of $\Lambda_{\rm uv}$, order by order in the expansion in powers of $\lambda$,  so that the physical quantities remain finite as $\Lambda_{\rm uv}\to \infty$. This procedure fixes the dependence of $m_{\rm b}$ on $\Lambda_{\rm uv}$ but leaves undetermined the finite part of $m_{\rm b}$. The latter is  fixed  by a renormalization condition, for instance
\beq\label{renormcond}
\Gamma^{(2)}(p=0)=m^2,
\eeq
where $m$ is the renormalized (or physical) mass, which we shall keep finite. (The massless case would require facing issues related to infrared divergences. Although these can be easily handled by the renormalization group techniques that we shall discuss later, their discussion would be an unnecessary distraction in this paper.)

\begin{figure}[htbp]
\begin{center}
\includegraphics[width=5cm]{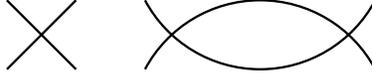}
\caption{The contributions to the kernel ${\cal I}(q,p)$ to orders $\lambda$ and $\lambda^2$. }\label{fig:kernel}
\end{center}
\end{figure}

This renormalization procedure can be generalized to the case of non perturbative calculations based on the  2PI effective action and $\Phi$-derivable approximations. Let us   recall that the central quantity in such approximations is $\Phi[G]$, the sum of the two-particle-irreducible ``skeleton'' diagrams, a  functional of the full propagator $G$. From $\Phi[G]$  one obtains the self-energy by functional differentiation (to within factors $(2\pi)^3)$:
\beq\label{PhiPi}
\Sigma(p)=2 \frac{\delta \Phi}{\delta G(p)}.
\eeq
This relation, together with Dyson's equation:
\beq\label{Dyson}
G^{-1}(p)=p^2+m_{\rm b}^2+\Sigma(p)\,,
\eeq
defines the physical propagator and self-energy in a self-consistent way. We shall refer to Eq.~(\ref{Dyson}), with $\Sigma[G]$ given by Eq.~(\ref{PhiPi}), as the ``gap equation''. A further differentiation of $\Phi[G]$ with respect to $G$ yields the two-particle-irreducible kernel
\beq\label{eq:Lambda}
{\cal I}(q,p)=2\frac{\delta \Sigma(p)}{\delta G(q)}=4\frac{\delta^2\Phi}{\delta G(q)\delta G(p)}={\cal I}(p,q)
\eeq
of a Bethe-Salpeter  type equation
\beq\label{BS1}
\Gamma^{(4)}(q,p) & = & {\cal I}(q,p)-\frac{1}{2}\int_l \Gamma^{(4)}(q,l)\,G^2(l)\,{\cal I}(l,p)\nonumber\\
& = & {\cal I}(q,p)-\frac{1}{2}\int_l {\cal I}(q,l)\,G^2(l)\,\Gamma^{(4)}(l,p)
\eeq
that allows the calculation of the four-point function $\Gamma^{(4)}(q,p)\equiv\Gamma^{(4)}(q,-q,p,-p)$: the quantity ${\cal I}(q,p)$ is the two-particle-irreducible contribution to $\Gamma^{(4)}(q,p)$ in one particular channel. If all skeletons are kept in $\Phi$, these relations  are exact. A $\Phi$-derivable
approximation \cite{Baym} is  obtained by selecting a class of skeletons in $\Phi$ and calculating $\Sigma$ and $\Gamma^{(4)}$ from the equations above. For instance, the 3-loop approximation to $\Phi$ is the following functional of $G$: 
\beq\label{Phi3loop}
\Phi[G]=\frac{\lambda}{8}\left(\int_q G(q)\right)^2-\frac{\lambda^2}{48}\int_p\int_q\int_l G(p)G(q)G(l)G(l+q+p),
\eeq
and the corresponding skeletons that contribute  to $\Sigma[G]$ and to ${\cal I}[G]$ in this approximation are displayed respectively  in  Figs.~\ref{fig:example} and \ref{fig:kernel}.

The equations~(\ref{PhiPi}-\ref{Phi3loop})  involve  integrals that would be divergent in the absence of the ultraviolet regulator. 
In three dimensions, and for any truncation of $\Phi$, the only divergences of the gap equation are again those of the ``tadpole'' and ``sunset" diagrams, and these divergences are independent of the momentum $p$.  It follows that the 2-point function can be made finite by absorbing these divergences into the bare mass, with the finite part determined as before by the renormalization condition:
\begin{equation}\label{eq:bm}
m^2=m_{\rm b}^2+\Sigma(p=0).
\end{equation}
The renormalized propagator is then given by 
\beq\label{renormprop}
G^{-1}(p)=p^2+m^2+\left[ \Sigma(p)-\Sigma(0)  \right].
\eeq
With $\Sigma$ given in terms of $G$ by Eq.~(\ref{PhiPi}), this equation (\ref{renormprop}) may also be viewed as the renormalized gap equation. 
Note  that now, the function $m_{\rm b}^2(m^2,\Lambda_{\rm uv},\lambda)$ in Eq.~(\ref{eq:bm}) is determined from the solution of a non-linear equation (rather than as an expansion in powers of $\lambda$, as in the perturbative case). However it is not needed: in practice, one can work directly with the renormalized propagator, with no reference to $m_{\rm b}$.\\

We now turn to the exact renormalization group, of which there exist  several variants  (for reviews see e.g.  \cite{Berges:2000ew}). We follow here  Ref.~\cite{Wetterich93}, and add to the original action $S[\varphi]$ a regulator   term $\Delta S_\kappa[\varphi]$ of the form
\beq
\Delta S_\kappa[\varphi]= \frac{1}{2}\int_q\: R_\kappa(q)\varphi(q)\varphi(-q),\qquad \int_q\equiv \int\frac{d^3 q}{(2\pi)^3}\,,
\eeq 
where the parameter $\kappa$ runs continuously from a  ``microscopic scale'' $\Lambda$ (to be specified below) down  to 0. The role of $\Delta S_\kappa$ is that of a mass term that suppresses the fluctuations with momenta lower than $\kappa$, while leaving unaffected those with momenta greater than  $\kappa$. This is usually achieved with a smooth cut-off function  $R_\kappa(q)$ such that $R_\kappa(q\ll \kappa)\rightarrow \kappa^2$ and $R_\kappa(q)$ goes rapidly to 0 as $q\simge \kappa$ (so that $\partial_\kappa R_\kappa(q)$ can play the role of an ultraviolet cut-off) in the flow equation. 

The effective action $\Gamma_\kappa[\phi ]$ associated to $S+\Delta S_\kappa$ obeys the exact flow equation \cite{Wetterich93} 
 \beq \label{eq:gammaflow}
\pk \Gamma_\kappa[\phi]=\frac{1}{2}\int_q\,\partial_\kappa R_\kappa(q)\,G_\kappa[q,-q;\phi],
\eeq
where $G_\kappa[q,-q;\phi]$ is the full propagator in the presence of the background field $\phi$:
\beq\label{eq:prop}
G_\kappa^{-1}[q,-q;\phi]=\Gamma_\kappa^{(2)}[q,-q;\phi]+R_\kappa(q)\,,
\eeq
with $\Gamma^{(2)}_\kappa[q,-q;\phi]$  the second functional derivative of $\Gamma_\kappa[\phi]$ with respect to $\phi(q)$, $\phi(-q)$.  The (renormalized) effective action $\Gamma[\phi]$ of the scalar field theory  is obtained as the solution (for an appropriate initial condition) of  Eq.~(\ref{eq:gammaflow}) 
for $\kappa\to 0$, at which point $R_\kappa(q)$ vanishes.

Field derivatives of Eq.~(\ref{eq:gammaflow}) lead to an infinite hierarchy of  flow equations for the proper vertices or $n$-point functions $\Gamma^{(n)}_\kp(p_1,\cdots,p_n)$. For instance the flow equation for the two-point function for vanishing field ($\phi=0$) reads
\beq\label{eq:2flow}
\pk \Gamma^{(2)}_\kappa(p)=-\frac{1}{2}\int_q\,\pk R_\kp(q)\,G_\kappa^2(q)\,\Gamma^{(4)}_\kp(q,p)\,,
\eeq
where we have denoted the two- and four-point functions at vanishing field by $\smash{\Gamma^{(2)}_\kp(p)\equiv\Gamma^{(2)}_\kp(p,-p)}$,  $\smash{\Gamma^{(4)}_\kp(q,p)\equiv\Gamma^{(4)}_\kp(q,-q,p,-p)}$ (as we did after Eq.~(\ref{BS1})), and $G_\kappa(q)\equiv G_\kappa[q,-q;\phi=0]$. Equation (\ref{eq:2flow}) gives the flow of the two-point function in terms of the latter and the four-point function. Similarly, the flow of the four-point function can be expressed in terms of the two-, four- and six-point functions. And so on: successive derivatives yield an infinite hierarchy of coupled flow equations for the $n$-point functions, whose solution requires a truncation of some form. We shall discuss soon a truncation that exploits the relations that hold in the 2PI formalism that has been summarized above.  But before we do that, we need to discuss further the initial conditions that need to be imposed on the flow in order to recover the results of standard field theoretical calculations. 

These initial conditions are commonly imposed at the microscopic scale $\kappa=\Lambda$ that we have mentioned earlier: this scale $\Lambda$ is the scale at which, on general grounds \cite{Berges:2000ew},  one expects  the effective action to take the form of the classical action of Eq.~(\ref{eactON}), with suitably defined parameters. There are however subtle issues related to the precise behavior of the $n$-point functions in the vicinity of the scale $\Lambda$, that need to be examined. We shall do that first within perturbation theory, focusing on the 2-point function $\Gamma_\kappa^{(2)}(p)$. We  anticipate that, when $\Lambda$ is large enough, $\Gamma^{(2)}_\Lambda(p)$ should be of the form $\Gamma^{(2)}_\Lambda(p)=p^2+m_\Lambda^2$, with $m_\Lambda$ independent of $p$, and we shall verify that it is indeed the case.

Let us recall that perturbation theory   can be recovered  from the flow equations by solving them recursively, taking the classical action as  initial condition. For instance, in leading order, the equation for the 2-point function reads
\beq\label{eq:2flowpert}
\pk \Gamma^{(2)}_\kappa(p)=-\frac{\lambda}{2}\int_q\,\pk R_\kp(q)\left( G^{(0)}_\kappa(q)\right)^2,
\eeq
where $G_\kappa^{(0)}(q)=\left(q^2+m_{\rm b}^2+R_\kappa(q)\right)^{-1}$. We can write this as
\beq\label{eq:2flowpert2}
\pk \Gamma^{(2)}_\kappa(p)=\frac{\lambda}{2}\,\partial_\kappa \int_q\,\left[ G^{(0)}_\kappa(q)-G^{(0)}(q)\right],
\eeq
where the subtracted term, which  makes the integral convergent, is nothing but the leading order counterterm, $\delta m^2_{\rm b}=-(\lambda/2)\int_q G^{(0)}(q)$ (with $m^2_{\rm b}\equiv m^2+\delta m_{\rm b}^2$). The integration of 
Eq.~(\ref{eq:2flowpert2})  is  immediate, and yields
\beq\label{Gamma2tadpole}
\Gamma_\kappa^{(2)}(p)=p^2+m^2+\left[\Sigma_\kappa^{\rm tad}-\Sigma^{\rm tad}\right],\qquad \Sigma_\kappa^{\rm tad}\equiv \frac{\lambda}{2} \int_q\, G^{(0)}_\kappa(q),
\eeq
where, as indicated,  $\Sigma_\kappa^{\rm tadpole}$ is the contribution of the tadpole diagram evaluated with the propagator $G^{(0)}_\kappa(q)$, and  we have used the renormalization condition at $\kappa=0$, Eq.~(\ref{renormcond}). 
Defining a running mass $m_\kappa$ by
\beq\label{runningmass}
m_\kappa^2=m^2+\left[\Sigma_\kappa(p=0)-\Sigma(p=0)\right],
\eeq
we can rewrite Eq.~(\ref{Gamma2tadpole}) as follows
\beq\label{Gamma2tadpole2}
\Gamma_\kappa^{(2)}(p)=p^2+m_\Lambda^2+\left[\Sigma_\kappa^{\rm tad}-\Sigma_\Lambda^{\rm tad}\right].
\eeq
This equation shows that one can indeed obtain $\Gamma_\kappa^{(2)}(p)$ by integrating the flow equation from the scale $\Lambda$ down to $\kappa$, with the initial condition $\Gamma^{(2)}_\Lambda(p)=p^2+m_\Lambda^2$. 
Note that one may, if one wishes, eliminate the ultraviolet cut-off, i.e., let $\Lambda_{\rm uv}\to\infty$, the resulting divergence of the self-energy canceling  when differences are taken, as e.g. in Eq.~(\ref{runningmass}). 
Also, at this order, one may replace $m_{\rm b}^2$ in $G_\kappa^{(0)}(q)$ by $m^2$, or $m_\Lambda^2$. Then one may completely ignore where $m_\Lambda$ is coming from and consider it as a parameter characterizing the initial condition, to be adjusted so as to satisfy the renormalization condition.  This property persists in higher orders of perturbation theory, and we shall verify it explicitly in second order. 
  
Consider then perturbation theory in second  order, and focus  on the sunset diagram, $\Sigma_\kappa^{\rm sun}(p)$, which is the only second order diagram that depends on the external momentum $p$. In order to manipulate only quantities that remain finite when $\Lambda_{\rm uv}$ is sent to infinity, we write the contribution of the sunset diagram to $\Gamma^{(2)}_\kp(p)$ as follows
\beq\label{eq:gap3l2}
\Sigma_\kappa^{\rm sun}(p)-\Sigma^{\rm sun}(0)=\left[\Sigma_\kappa^{\rm sun}(0)-\Sigma^{\rm sun}(0)\right]+\left[\Sigma_\kappa^{\rm sun}(p)-\Sigma_\kappa^{\rm sun}(0)\right].\eeq
The first term within brakets in the right-hand side contributes a correction to the running mass $m_\kappa^2$ (see Eq.~(\ref{runningmass})). The second term is a function, $\Delta^{\rm sun}_\kappa(p)\equiv \Sigma_\kappa^{\rm sun}(p)-\Sigma_\kappa^{\rm sun}(0)$, that vanishes as $p\to 0$:
\beq\label{eq:gap3l1}
\Delta_\kappa^{\rm sun}(p)=-\frac{\lambda^2}{6}\int_q\int_l \Big[G^{(0)}_\kp(q)G^{(0)}_\kp(l)G^{(0)}_\kp(l+q+p)-G_\kappa^{(0)}(q)G_\kappa^{(0)}(l)G_\kappa^{(0)}(l+q)\Big].
\eeq
 In order to verify that the initial condition of the flow remains of the form $\Gamma^{(2)}_\Lambda(p)=p^2+m_\Lambda^2$, we need to  study the behavior of  $\Delta_\kappa^{\rm sun}(p)$ and show that it vanishes for $\kappa=\Lambda\gg p$ large. To do so, we  rewrite Eq.~(\ref{eq:gap3l1}) as follows
 \beq\label{Gamma2tilde}
\Delta_\Lambda^{\rm sun}(p)=-\frac{\lambda^2}{6}\int_{\tilde q}\int_{\tilde l} \Big[\tilde G^{(0)}_\Lambda(\tilde q)\tilde G^{(0)}_\Lambda(\tilde l)\tilde G^{(0)}_\Lambda(\tilde l+\tilde q+\tilde p)-\tilde G_\Lambda^{(0)}(\tilde q)\tilde G_\Lambda^{(0)}(\tilde l)\tilde G_\Lambda^{(0)}(\tilde l+\tilde q)\Big],
\eeq
where  we have rescaled the integration variables by $\Lambda$, e.g. $\tilde q=q/\Lambda$, we have set $R_\Lambda(q)=\Lambda^2 r(\tilde q)$, and 
$
\tilde G^{(0)}_\Lambda(\tilde q)=\left( \tilde q^2+r(\tilde q)+m^2/\Lambda^2\right)^{-1}.
$
Clearly, the  expression (\ref{Gamma2tilde})  vanishes at least as fast as $p^2/\Lambda^2$ (one assumes $r$ to be a smooth function of its argument, respecting rotational symmetry). Thus, at order $\lambda^2$, the initial condition retains the same form,  $\Gamma^{(2)}_\Lambda(p)=p^2+m_\Lambda^2$:  the momentum dependent contributions to $\Delta^{\rm sun}_\kappa(p)$ are suppressed at large values of $\Lambda$ by powers of $p/\Lambda$, leaving eventually only the tree-level contribution $\sim p^2$ in $\Gamma^{(2)}_\Lambda(p)$. It is not difficult to extend this result to all orders: in  the present 3-dimensional theory power counting is enough to do so.

 In fact, a simple  dimensional analysis allows us to exhibit the large $\kappa$ behavior of the $n$-point functions. By considering successively the flow equations of increasing orders, it is not difficult to show that,  at large values of $\kappa$ (larger than any external momentum), and at leading order, the flow equations for the $n$-point functions admit the following self-consistent solution: 
$
\Gamma^{(2)}_\kp(p)\equiv p^2+m^2_\kappa$, with $m^2_\kappa\sim  \lambda\kappa$,   $\Gamma^{(4)}_\kappa\sim \lambda$ (to within a correction $\sim \lambda^2/\kappa$), and $ \Gamma^{(2n\ge 3)}_\kappa\sim \lambda^n/\kappa^{2n-3}$. To obtain this, we have taken into account that the loop momentum $q$ in the flow equation (\ref{eq:gammaflow}) is bounded by $\kappa$ (because  $\partial_\kappa R_\kappa(q)$ plays the role of an ultraviolet  cut-off at scale $\kappa$). We also used the condition $\Gamma^{(4)}_{\kappa\to\infty}=\lambda$,  as well as  the absence of tree level couplings for the higher $n$-point functions. This behavior of the $n$-point functions is consistent both with the expectation that the effective action reduces to the classical action for large values of $\kappa$, and with perturbation theory: the term $\lambda^n/\kappa^{2n-3}$ which drives the flow of $ \Gamma_\kappa^{(2n)}$ coincides indeed with the leading perturbative contribution to $ \Gamma_\kappa^{(2n)}$.  Note that the  scaling  analysis that we have just presented yields the dominant (``divergent'')  contribution  to the 2-point function, but a more refined analysis is needed to get the subleading terms (in particular the ``finite part'' of $\Gamma^{(2)}_\kp(p)$): this is what we have done explicitly earlier by examining explicitly the first two orders of perturbation theory.

We are now ready to discuss a specific truncation of the flow equations, which consists in imposing  the relation (\ref{BS1}) for selected skeleton diagrams. Since ${\cal I}$ is a functional of the 2-point function, this allows us indeed to close the flow hierarchy. We obtain:
\beq
\pk \Gamma^{(2)}_\kappa(p)= & - & \frac{1}{2}\int_q \,\pk R_\kappa(q)\,G^2_\kp(q)\,\Gamma^{(4)}_\kp(q,p)\label{eq:flow2}\,,\\
\Gamma^{(4)}_\kp(q,p)={\cal I}_\kp(q,p) & - & \frac{1}{2}\int_l\,\Gamma^{(4)}_\kp(q,l)\,G^2_\kp(l)\,{\cal I}_\kappa(l,p)\,,\label{eq:BS2}
\eeq
where the subscript $\kappa$ on $\mathcal{I}_\kappa$ means that the functional derivative defining the kernel $\mathcal{I}$ (see Eq.~(\ref{eq:Lambda})) is to be evaluated for $\smash{G=G_\kappa}$, with $G_\kappa$ related to $\Gamma^{(2)}_\kappa$ by Eq.~(\ref{eq:prop}). Note that Eq.~(\ref{eq:BS2}) is ultraviolet finite, but in contrast to what happens in Eq.~(\ref{eq:flow2}), this is not due to the derivative of the regulator, namely to the term $\partial_\kappa R_\kappa(q)$ in Eq.~(\ref{eq:flow2}), but results here from the property of the 3-dimensional integral in Eq.~(\ref{eq:BS2}), and can be verified by simple power counting. This property also holds in the four-dimensional case  \cite{BPR}, although the proof there requires more work.

One nice feature of this truncation scheme is that it is systematically improvable, by adding more skeletons to $\Phi$: if all skeletons are included, the solution of the coupled system of equations (\ref{eq:flow2}-\ref{eq:BS2}) provides the exact 2-point function as well as the exact 4-point function for a particular configuration of the external momenta. A second attractive feature is that it preserves the property of the flow of being a total derivative with respect to the parameter $\kappa$. To see that,  let us plug Eq.~(\ref{eq:BS2}) into Eq.~(\ref{eq:flow2}). We obtain:
\beq\label{eq:exact1}
\partial_\kp\Gamma^{(2)}_\kp(p)= & - &\frac{1}{2}\int_q \,\pk R_\kp(q)\,G^2_\kp(q)\,{\cal I}_\kp(q,p)\nonumber\\
& + & \frac{1}{4}\int_l\int_q \,\pk R_\kp(q)\,G_\kappa^2(q)\,\Gamma^{(4)}_\kp(q,l)\,G^2_\kp(l)\,{\cal I}_\kp(l,p)\,.
\eeq
Using Eq.~(\ref{eq:flow2}) in the second line and renaming the dummy variable $l$ by $ q$, we arrive at:
\beq\label{eq:exact0}
\pk\Gamma^{(2)}_\kp(p)=-\frac{1}{2}\int_q \,(\pk R_\kp(q)+\pk\Gamma^{(2)}_\kp(q))\,G^2_\kp(q)\,{\cal I}_\kp(q,p)=\frac{1}{2}\int_q\,\pk G_\kp(q)\,{\cal I}_\kp(q,p)\,.
\eeq
Finally, using Eq.~(\ref{eq:Lambda}), we obtain:
\beq\label{eq:exact}
\pk\Gamma^{(2)}_\kp(p)=\int_q\pk G_\kp(q)\,\left.\frac{2\delta^2\Phi}{\delta G(q)\delta G(p)}\right|_{G_\kappa}=\pk\Sigma_\kappa(p),
\eeq
where, in  the last step, we have used the fact that the $\kappa$-dependence of $\Phi$ originates entirely from the $\kappa$-dependence of the propagator, and  $\Sigma_\kappa\equiv \Sigma[G_\kappa]$, with $\Sigma[G]$ given by Eq.~(\ref{PhiPi}). As announced, the truncated flow is an exact derivative with respect to $\kp$. This is a unique property of this truncation, that is not shared by most other popular truncations of the exact RG (with the noticeable exceptions of the perturbative expansion discussed above, and the large $N$ approximation, see e.g. \cite{Blaizot:2008xx}. A similar property of the flow equation, that of being an exact derivative, was also obtained in the off-equilibrium context in Ref.~\cite{Gasenzer:2008zz}). Since it is a total derivative, the flow can be formally integrated out and its general solution written as
\beq\label{eq:general_sol}
G^{-1}_\kp(p)=p^2+m^2+\gamma(p)+R_\kappa(p)+[\Sigma_\kp(p)-\Sigma(0)]\,.
\eeq
The function $\gamma(p)$ is an arbitrary integration ``constant'' (that may a priori depend on $p$, with however  $\gamma(0)=0$ so as to fulfill the renormalization condition). Equation (\ref{eq:general_sol}) may be viewed as a gap equation (see Eq.~(\ref{renormprop})) whose solution is equivalent to the solution of the flow equation within the chosen truncation (defined by a selection of skeletons contributing to $\Phi$). Equivalently, it performs a resummation of  all the  Feynman diagrams  that are generated from the corresponding skeletons, in a theory with action $S+\Delta S_\kappa+\frac{1}{2}\int_q\gamma(q)\varphi(q)\varphi(-q)$. This remark makes it clear that we have to choose $\gamma(p)=0$ (and not only $\gamma(0)=0$) if one wants the corresponding diagrams to be those of the $\varphi^4$ theory: this is equivalent to say that all the momentum dependence of $\Gamma^{(2)}(p)$, beyond the tree-level $p^2$ contribution, comes entirely from the self-energy diagrams, which removes all ambiguity in the initial condition.  Because the solution of the gap equation corresponds to an exact resummation of selected Feynman diagrams, at the end of the flow where $\kappa=0$ and the regulator vanishes, the final result is rigorously independent of the choice of the regulator.  (Of course, the values of $m_\Lambda$ may differ for various choices of regulators, but these differences can be explicitly calculated.) We should mention here a possible ambiguity in the correspondence between the flow equation and the gap equation: the flow equation has a unique solution for a given initial condition, while the non-linear gap equation may have several solutions. However, leaving aside possible artifacts of approximations, this situation is expected to occur only in cases of symmetry breaking, a situation that will not be discussed here.

All these properties of the 2PI truncation may look at first disappointing from the point of view of the flow equations: indeed, all what the flow does in this particular truncation is solving the 2PI equations! However, there is certainly interest in establishing direct connections between non trivial non-perturbative approximations. In particular, because the 2PI truncations lead to flow equations that are exact derivatives, they could be used to test other approximations, such as the derivative expansion, the vertex expansion  or the scheme proposed in Ref.~\cite{BMW}. Besides, from the point of view of the 2PI formalism, there is a practical advantage in reformulating the gap equation as a flow equation:  this is because initial value problems are in general easier to solve than non linear gap equations. We shall return to this practical aspect at the end of this letter. 

Because, in the 2PI truncation,  the behavior of the 4-point function is not controlled by the regulator term $\partial_\kappa R_\kappa$, as we have already emphasized, we need to reconsider carefully the initial conditions of the flow.  From Eq.~(\ref{eq:general_sol})  (in which we set $\gamma(p)=0$), we  deduce 
\beq\label{eq:init_simple}
\Gamma^{(2)}_\Lambda(p)=p^2+m^2+[\Sigma_\Lambda(p)-\Sigma(0)]=p^2+m_\Lambda^2+\Delta_\Lambda(p),
\eeq
where we have generalized the notation used in Eq.~(\ref{eq:gap3l1}), namely $\Delta_\kappa(p)=\Sigma_\kappa(p)-\Sigma_\kappa(0)$. 
We want to verify that $\Delta_\Lambda(p)$ goes to $0$ when $\Lambda\gg p$. We shall do that by examining  the successive terms in the loop expansion of $\Phi$. Since at order two-loop, $\Sigma$ is independent of $p$, the first non trivial term to consider is of order  three-loop. Its expression is similar to the perturbative one in Eq.~(\ref{Gamma2tilde}), with $m_\Lambda^2$  replaced by the full momentum function $m_\Lambda^2+\Delta_\Lambda(q)$. We get
\beq\label{SigmaLambda3l}
\Delta_\Lambda^{3\ell} (p)=-\frac{\lambda^2}{6}\!\int_{\tilde l}\!\int_{\tilde q}\,
\tilde G_\Lambda(\tilde l)
\, \tilde G_\Lambda(\tilde q) \left[ \tilde G_\Lambda(\tilde l+\tilde q+\tilde p)- \tilde G_\Lambda(\tilde l+\tilde q)\right],
\eeq
with 
\beq
 \tilde G_\Lambda(\tilde q) \equiv \frac{1}{\tilde q^2+r(\tilde q)\!+\! \tilde m_\Lambda^2\!+\! \tilde \Delta_\Lambda(\tilde q)},
 \qquad \tilde m_\Lambda^2\equiv \frac{m_\Lambda^2}{\Lambda^2}, \qquad \tilde \Delta_\Lambda(\tilde q)\equiv \frac{\Delta_\Lambda(q)}{\Lambda^2}
 \eeq
 and we have used the same scaled variables as in Eq.~(\ref{Gamma2tilde}).  To proceed further, we  assume that, at large $\Lambda$ ($\Lambda\gg\lambda$), $m^2_\Lambda\sim \lambda\Lambda$ and $\Delta_\Lambda(q)\sim \lambda^2  d(q/\Lambda)$ where the dimensionless function $d(q/\Lambda)$ grows slower than $ q^2$ when $ q\to \infty$. These assumptions, which we justify later, enable us to perform a Taylor expansion in powers of $\tilde p$ in Eq.~(\ref{SigmaLambda3l}) and show that  $\Delta_\Lambda^{3\ell} (p)\to 0$  as $p^2/\Lambda^2\to 0$ (recall that $r$ is assumed to be a smooth function of its argument, respecting rotational symmetry). Beyond the three-loop $\Phi$-derivable approximation, the proof follows form  simple power counting, with  the $l$-loop contributions to $\Sigma_\Lambda (p)$ being of order $\lambda^2(\lambda/\Lambda)^{l-2}$, and the momentum corrections analytic in $p^2/\Lambda^2$. \\

In order to verify the self-consistency of our assumption, we shall look first at the flow of the running mass, $m_\kappa^2$, at large $\kappa$ ($\kappa\gg \lambda$). In fact, since this provides a nice illustration of how the flow equations work within the present truncation, we shall push the calculation beyond what is strictly needed to verify the assumption that $m_\kappa\sim\lambda\kappa$ at large $\kappa$. From  Eq.~(\ref{eq:flow2}) we get
\beq\label{eq:flow2b}
\partial_\kappa m^2_\kappa=-\int_{\tilde q} \frac{ s(\tilde q)}{\left({\tilde q}^2+r(\tilde q)+\tilde m^2_\kappa+\tilde \Delta_\kappa( \tilde q)\right)^2}\;\Gamma^{(4)}_\kappa(q,0)\,,
\eeq
with, as earlier,  $R_\kappa(q)=\kappa^2r(\tilde q)$, $\pk R_\kp(q)=2\kp s(\tilde q)$ and  $\tilde q\equiv q/\kappa$. Using a similar notation, and our assumptions about $m_\kappa$ and $\Delta_\kappa(q)$, it is not difficult to show that the leading behavior of the kernel ${\cal I}_\kp( q,p)$ in Eq.~(\ref{eq:BS2}) is of the form
\beq\label{calI2loop}
{\cal I}_\kp( q,p)&=&\lambda-\frac{\lambda^2}{\kappa}\int_{\tilde l}\,\frac{1}{\tilde l^2+r(\tilde l)+m_\kappa^2+\tilde\Delta_\kappa(\tilde l)}\:\frac{1}{\tilde k^2+r(\tilde k)+\tilde m_\kappa^2+\tilde\Delta_\kappa(\tilde k)}\nonumber\\
 &=& \lambda-\frac{\lambda^2}{\kappa} b(\tilde q+\tilde p)+\mathcal{O}\left(\frac{\lambda^3}{\kappa^2}\right),
\eeq
where $k\equiv l+p+q$, and 
\beq
b(\tilde q)\equiv \int_{\tilde l} \frac{1}{\tilde l^2+r(\tilde l)}\:\frac{1}{(\tilde l+\tilde q)^2+r(\tilde l+\tilde q)}.
\eeq
 To the same accuracy, one finds from Eq.~(\ref{eq:BS2}):
\begin{eqnarray}\label{eq:exp}
\Gamma^{(4)}_\kappa(q,p) & = & {\cal I}_\kappa( q,p)-\frac{1}{2\kappa}\int_{\tilde l} {\cal I}_\kappa(q, l)\,\frac{1}{\left(\tilde l^2+r(\tilde l)+\frac{\Delta_\kappa( l)}{\kappa^2}\right)^2}\,\Gamma^{(4)}_\kp( l,p)\nonumber\\
& = & \lambda-\frac{\lambda^2}{2\kappa}\Big[b(0)+2b(\tilde q+\tilde p)\Big]+\mathcal{O}\left(\frac{\lambda^3}{\kappa^2}\right).
\end{eqnarray}
Note that  the three channels contribute to $\Gamma^{(4)}_\kappa(q,p)$ at order $\lambda^2$, two are included in the second order contribution to the kernel ${\cal I}_\kp( q,p)$, one is generated by the integral equation for $\Gamma^{(4)}_\kappa(q,p)$ (the last term in the first line of Eq.~(\ref{eq:exp}), with ${\cal I}$ and $\Gamma^{(4)}$ replaced by their leading order contribution, $\lambda$). 
Using this expression of $\Gamma^{(4)}_\kappa(q,p)$ in Eq.~(\ref{eq:flow2b}), one finds, after a simple calculation:
\begin{equation}\label{eq:abc}
\partial_\kp m^2_\kp=-\lambda A +\frac{\lambda^2}{\kp} B +\mathcal{O}\left(\frac{\lambda^3}{\kappa^2}\right), \quad A\equiv \int_{\tilde q}\frac{s(\tilde q)}{(\tilde q^2+r(\tilde q) )^2}\,, \quad B\equiv \int_{\tilde q}\frac{b(\tilde q)s(\tilde q)}{(\tilde q^2+r(\tilde q))^2}.
\end{equation}
Note that the term $b(0)$ in the expression (\ref{eq:exp}) of  $\Gamma^{(4)}_\kappa(q,p)$ has cancelled against a corresponding contribution that originates from expanding the denominator in Eq.~(\ref{eq:flow2b}) in powers of $m^2_\kappa/\kappa^2$, and keeping in this expansion only the leading term $m^2_\kappa/\kappa^2=-\lambda A/\kappa$. This cancellation, which results from the identity (a similar identity exists in $d$-dimensions)
\beq
\int_{\tilde q}\frac{1}{(\tilde q^2+r(\tilde q))^2}=4\int_{\tilde q}\frac{s(\tilde q)}{(\tilde q^2+r(\tilde q))^3}\,,
\eeq
is to be expected: only the two channels included in ${\cal I}_\kp( q,p)$ are responsible for the logarithmic behavior.  The flow of $m^2_\kp$ for $\kappa\gg p$, as given by Eq.~(\ref{eq:abc}) fixes the $\Lambda$ dependence of $m_\Lambda^2$, beyond the leading order needed in our previous proof: 
$ m^2_\Lambda=-\lambda A \Lambda +\lambda^2 B\ln (\Lambda/\lambda)+\cdots$. 

In order to verify our assumption on $\Delta_\kappa(p)$ at large $\kappa$, and large $p$, the most convenient is to return to Eq.~(\ref{SigmaLambda3l}). It is then not difficult to show that the dominant behavior at large $p$ ($p\gg\Lambda$) is  $\Delta_\Lambda(p)\sim \lambda^2 \ln (p^2/\Lambda^2)$. This indeed satisfies our assumption.

The  results presented in this letter could be used to simplify the solution of $\Phi$-derivable approximations. Instead of solving the  non-linear integral equation (\ref{Dyson}), we can consider the initial value problem (\ref{eq:flow2}) coupled to the linear integral equation (\ref{eq:BS2}). In fact, this can be  further simplifed. Rewrite  indeed the truncated flow, Eqs.~(\ref{eq:flow2})-(\ref{eq:BS2}), as $\pk\Gamma^{(2)}_\kp(p)=F_\kp(p)$ with
\beq
F_\kp(p)= {\cal J}_\kp(p) & - & \frac{1}{2}\int_q\,F_\kp(q)\,G_\kappa^2(q)\,{\cal 
I}_\kp(q,p)\,,\label{eq:Fk}\\
 {\cal J}_\kp(p)= & - & \frac{1}{2}\int_q\,\pk R_\kp(q)\,G_\kp^2(q)\,{\cal 
I}_\kappa(q,p)\,.\label{eq:tildeI}
\eeq
At each integration step in $\kappa$, one evaluates ${\cal J}_\kp(p)$ from Eq.~(\ref{eq:tildeI}) and solves the linear integral equation (\ref{eq:Fk}) in order to determine $F_\kappa(p)$. One benefit  is that the function to be determined in Eq.~(\ref{eq:Fk}) depends on a single momentum, whereas $\Gamma_\kappa^{(4)}(q,p)$ in Eq.~(\ref{eq:BS2})  depends on two momenta.

As a final remark, let us mention that we can further exploit the  freedom in the way we may implement the regulator, departing in doing so from the traditional RG approach. Recall that the solution of the 2PI truncated flow, with the initial condition discussed above, is identical to the solution of the renormalized gap equation (\ref{eq:general_sol}) (with $\gamma(p)=0$). Consider now the following  gap equation
\beq\label{eq:gapnew}
\hat G^{(-1)}_\kp(p)=p^2+m^2+R_\kappa(p)+\hat \Delta_\kappa(p) , \qquad \hat \Delta_\kappa(p)\equiv \hat\Sigma_\kappa(p)-\hat \Sigma_\kappa(0),
\eeq
with $\hat \Sigma_\kp=\Sigma[\hat G_\kp]$. Clearly,  $\Gamma^{(2)}_\kappa(p)$ and $\hat\Gamma^{(2)}_\kappa(p)=\hat G^{(-1)}_\kp(p)-R_\kappa(p)$ coincide for $\kappa=0$ (assuming unicity of the solution). Moreover, an analysis similar to the one performed above reveals that $\hat\Delta_\kappa(p)$ is suppressed for large $\kp$. Thus the initial condition for $\hat\Gamma^{(2)}_\kp(p)$ involves directly the physical mass, rather than $m_\Lambda$: $\hat\Gamma^{(2)}_\Lambda(p)=p^2+m^2$.  No fine tuning of $m_\Lambda$ needs to be done. This mirrors the fact, that in the standard formulation of $\Phi$-derivable approximations, it is also possible to rewrite the gap equation explicitely in terms of the renormalized mass, as observed after  Eq.~(\ref{renormprop}). 

In order to derive the flow equation for  $\hat\Gamma^{(2)}_\kp(p)$, we first notice that the flow equation for $\Gamma^{(2)}_\kp(p)$ could have been obtained from Eq.~(\ref{eq:general_sol}) by performing all the steps in Eqs.~(\ref{eq:exact}-\ref{eq:exact1}) backwards.  If we apply this strategy to Eq.~(\ref{eq:gapnew}), we obtain:
\beq
\pk\hat\Gamma^{(2)}_\kp(p) & - & \frac{1}{2}\int_q\pk R_\kp(q)\,\hat G_\kp^2(q)\,\hat\Gamma^{(4)}_\kp(q,p)\,,\label{flottildegamma2}\\
\hat\Gamma^{(4)}_\kp(q,p)=\hat{\cal I}_\kp(q,p) & - & \frac{1}{2}\int_l\hat\Gamma^{(4)}_\kp(q,l)\,\hat G_\kp^2(l)\,\hat{\cal I}_\kp(l,p)\,.\label{tildegamma4}
\eeq
The difference with the standard flow equation is that the kernel is given by $\hat {\cal I}_\kp(q,p)={\cal I}_\kp(q,p)-{\cal I}_\kp(q,0)$, which is no longer symmetric. One can also rewrite these equations in a form similar to Eqs.~(\ref{eq:Fk}-\ref{eq:tildeI}). The important, and unusual,  aspect of these equations is that they describe a flow at constant mass.

\end{document}